\documentclass[11pt]{article}
\usepackage{ifthen}
\usepackage{mdwlist}
\usepackage{amsmath,amssymb,amsfonts,amsthm}
\usepackage{bm}
\usepackage{bbm}
\usepackage[colorlinks=true,linkcolor=blue,citecolor=blue,urlcolor=blue]{hyperref}
\usepackage{enumitem}
\usepackage{graphicx}
\usepackage{xspace}
\usepackage{verbatim}
\usepackage{algorithm}
\usepackage{algpseudocode}
\usepackage[margin=1in]{geometry}
\usepackage{color}
\usepackage{thm-restate}
\usepackage{latexsym}
\usepackage{epsfig}
\usepackage[symbol]{footmisc}
\usepackage[colorinlistoftodos,textsize=scriptsize]{todonotes}
\def\withcolors{1}
\def\withnotes{1}
\def\eps{\ve}
\renewcommand{\epsilon}{\ve}
\def\ve{\varepsilon}

\newcommand{\pr}[2][]{\mathrm{Pr}\ifthenelse{\not\equal{}{#1}}{_{#1}}{}\!\left[#2\right]}

\newcommand{\dtv}{d_{\mathrm {TV}}}

\newtheorem{theorem}{Theorem}

\newtheorem{lemma}[theorem]{Lemma}

\newtheorem{corollary}[theorem]{Corollary}

\newtheorem{definition}[theorem]{Definition}

\numberwithin{theorem}{section} 
\numberwithin{nontheorem}{section} 
\numberwithin{proposition}{section} 
\numberwithin{observation}{section} 
\numberwithin{remark}{section} 
\numberwithin{fact}{section} 
\numberwithin{lemma}{section} 
\numberwithin{claim}{section} 
\numberwithin{corollary}{section} 
\numberwithin{case}{section} 
\numberwithin{dfn}{section} 
\numberwithin{definition}{section} 
\numberwithin{question}{section} 
\numberwithin{openquestion}{section} 
\numberwithin{res}{section}

\ifnum\withcolors=1
  \newcommand{\gcolor}[1]{{\color{red}#1}} 
  
  \newcommand{\hcolor}[1]{{\color{blue}#1}} 
\else

  \newcommand{\gcolor}[1]{{#1}}
  
  \newcommand{\hcolor}[1]{{#1}} 
\fi

\ifnum\withnotes=1
  \newcommand{\gnote}[1]{\par\gcolor{\textbf{G: }\sf #1}} 
  \newcommand{\gfootnote}[1]{\footnote{{\bf \gcolor{Gautam}}: {#1}}}
  
  \newcommand{\hnote}[1]{ \hcolor{\textbf{H: }\sf #1}} 
  \newcommand{\snote}[1]{ \hcolor{\textbf{SW: }\sf #1}} 

\else
  \newcommand{\gnote}[1]{}
  \newcommand{\hnote}[1]{}
    \newcommand{\snote}[1]{}

  \newcommand{\gfootnote}[1]{}

\fi
\newcommand{\ignore}[1]{\leavevmode\unskip} 
\newcommand{\gmargin}[1]{}

\DeclareMathOperator*{\argmin}{argmin}

\usepackage{mathtools}

\title{Sorting and Selection in Rounds with Adversarial Comparisons}

\author{
Chris Trevisan\thanks{University of Waterloo. {\tt chris.trevisan@uwaterloo.ca}.}
}
\date{}
\begin{document}
\maketitle\gmargin{The title seems fine. I might also consider ``Parallel Sorting and Selection with Adversarial Comparisons.'' But probably worth sticking as is.}

\begin{abstract}
  We continue the study of selection and sorting of $n$ numbers under the adversarial comparator model, where comparisons can be adversarially tampered with if the arguments are sufficiently close.

  We derive a randomized sorting algorithm that does $O(n \log^2 n)$ comparisons and gives a correct answer with high probability, addressing an open problem of Ajtai, Feldman, Hassadim, and Nelson \cite{AFHN15}. 
  Our algorithm also implies a selection algorithm that does $O(n \log n)$ comparisons and gives a correct answer with high probability. Both of these results are a $\log$ factor away from the naive lower bound. \cite{AFHN15} shows an $\Omega(n^{1+\eps})$ lower bound for both sorting and selection in the deterministic case, so our results also prove a discrepancy between what is possible with deterministic and randomized algorithms in this setting.

  We also consider both sorting and selection in rounds, exploring the tradeoff between accuracy, number of comparisons, and number of rounds. 
  Using results from sorting networks, we give general algorithms for sorting in $d$ rounds where the number of comparisons increases with $d$ and the accuracy decreases with $d$. Using these algorithms, we derive selection algorithms in $d+O(\log d)$ rounds that use the same number of comparisons as the corresponding sorting algorithm, but have a constant accuracy. Notably, this gives selection algorithms in $d$ rounds that use $n^{1 + o(1)}$ comparisons and have constant accuracy for all $d = \omega(1)$, which still beats the deterministic lower bound of $\Omega(n^{1+\eps})$.

\end{abstract}

\section{Introduction}
Comparison-based sorting and selection are two of the most well-studied computational problems, with applications in all aspects of computing. Often, these problems are studied with the goal to minimize the number of comparisons needed. Classical results show that sorting takes $\Theta(n \log n)$ comparisons \cite{MERGESORT} and selection takes $\Theta(n)$ comparisons \cite{SELECTION}.

Comparison-based sorting and selection have also been extensively studied in the parallel case. The round-based model of parallelism we consider was introduced by Valiant \cite{VAL} for comparison-based problems, where groups of comparisons are done in rounds of interaction. There followed a long line of research in parallel sorting and selection \cite{SR5,SR6,SR3,BT83,SR4,SR1,SR2,SR7}. Particularly similar to the problems studied in this paper are parallel sorting with limited closure \cite{BT83,Alo86} and sorting networks of arity $k$ with low depth \cite{SN1, SN2, SN4, SN3, SN5, SNLEC, SN6, SN8, SN9, SN7, DKP22}, the latter of which we will use to derive our general sorting algorithms.

However, often it is not possible to guarantee comparisons are completely precise. For example, when ranking chess players, or comparing job applicants. Due to this, the problems of sorting and selection with imprecise comparators have also been widely considered.

Depending on the model, the manner in which the comparator is imprecise differs. The adversarial comparison model we will study was introduced in \cite{AFHN15}. If the values being compared differ by more than some threshold $\delta$, the comparison is correct, otherwise the result of the comparison can be chosen arbitrarily by an adversary. There are two adversary models that have been considered (as described in \cite{AFJ18}): the \emph{non-adaptive} model, where all of the comparisons must be predetermined by the adversary before the algorithm is run, and the \emph{adaptive} model, where the comparisons can be chosen by the adversary at the time they are queried, possibly depending on the previous queries made by the algorithm. In this paper, we focus entirely on the adaptive model, and as our results are all upper bounds, all of our results imply equivalent results for the easier non-adaptive model. By scaling, we assume $\delta = 1$.

Since the comparisons are imprecise, it is impossible to always determine the correct result, so algorithms in this setting instead strive to achieve a small approximation factor, which measures how far away the returned solution is from the correct one. More precisely, we say an ordering $Y$ of a set $X$ is a $k$-approximate sorting if all inversions differ in value by at most $k$. In their original paper, Ajtai, Feldman, Hassidim, and Nelson \cite{AFHN15} give deterministic $k$-approximate algorithms for sorting and selection that use $O(4^k\cdot n^{1+1/2^{k-1}})$ and $O(2^k\cdot n^{1+1/2^{k-1}})$ comparisons respectively. They also give a lower bound of $\Omega(n^{1+1/2^{k-1}})$ for deterministic $k$-approximate sorting and selection. Special consideration is given to the case of selecting the maximum element, for which they give a randomized algorithm that uses $O(n)$ comparisons and returns a $3$-approximation with probability $1-n^{-r}$.
This maximum selection result was then improved by Acharya, Falahatgar, Jafarpour, Orlitsky, and Suresh \cite{AFJ18} who gave a randomized algorithm that uses $O(n \log \frac{1}{\epsilon})$ comparisons and returns a $2$-approximation with probability $1-\epsilon$. The study of these problems in the parallel setting was introduced by Gopi, Kamath, Kulkarni, Nikolov, Wu, and Zhang \cite{GKK20}, where they gave a randomized $d$-round algorithm that uses $O(n^{1+\frac{1}{2^d-1}}d)$ comparisons and returns a $3$-approximate maximum with probability $0.9$. This raises the following questions: can \emph{randomized} algorithms yield an improvement in sorting and general selection? How many comparisons are required to do sorting and general selection in $d$ rounds?

\subsection{Results, Techniques, and Discussion}
To describe our results, we more formally define the model and the problems of approximate sorting and selection.

\begin{definition}
  Suppose we are given $n$ items $x_1, \ldots, x_n$ with unknown real values. An adversarial comparator $C$ is a function that takes two items $x_i$ and $x_j$ and returns $\max\{x_i, x_j\}$ if $|x_i - x_j| > 1$ and $x_i$ or $x_j$ adversarially otherwise. 
\end{definition}

Throughout this paper, we assume the \emph{adaptive adversary} model \cite{AFJ18}, where the adversarial comparisons may depend on previous queries made by the algorithm.

We first define a notion of $k$-approximate sorting, in which inversions may only occur between values that differ by at most $k$.

\begin{definition}
    We say $x_i \geq_k x_j$ if $x_i \geq x_j - k$. For sets of items $Y,Z$, we say $Y \geq_k Z$ if $x_i \geq_k x_j$ for all $x_i \in Y$, $x_j \in Z$. We say some ordering $x_1,\ldots,x_n$ of items in a set $X$ is a $k$-approximate sorting if $x_j \geq_k x_i$ for all $j > i$. Equivalently, for any pair $x_i, x_j$ in the wrong order, they must differ by at most $k$.
\end{definition}

This leads to a notion of approximate $i$-selection, in which the result must be the $i$-th element of some approximate sorting.

\begin{definition}
    We say an item $x^*$ in a set $X$ is a $k$-approximate $i$-selection if there exists a $k$-approximate sorting $x_1,\ldots,x_n$ of $X$ such that $x_i = x^*$. 
\end{definition}

We show that this definition is equivalent to the result differing from the ``actual'' $i$-th smallest element by at most $k$.

\begin{restatable}{lemma}{lb}\label{lemma:KAPPROX}
    An item $x_j$ is a $k$-approximate $i$-selection if and only if $|x_j - x_i| \leq k$ where $x_i$ is the actual $i$-th smallest element of $X$.
\end{restatable}

We proceed with our results. We begin in the non-parallel setting (although our algorithms still have good round guarantees). We provide the following near-optimal approximate sorting algorithm.

\begin{restatable}{theorem}{lb}\label{thm:RSort}
There exists a randomized algorithm that takes $O(n \log^2 n)$ comparisons, uses $O(\log n)$ parallel rounds, and returns a $4$-approximate sorting with probability $> 1 - \frac{1}{n^2}$.
\end{restatable}

Since any approximate sorting algorithm must be able to correctly sort any list of numbers after scaling, such an algorithm must take $\Omega(n \log n)$ comparisons by the well-known lower bound. 
Thus, this result is a $\log$ factor away from optimal. Note that no algorithm can give better than a $2$-approximation, as the adversary can force $0 > 1 > 2 > 0$, which can make $0,1,2$ indistinguishable \cite{AFJ18}. The best prior result is of \cite{AFJ18}, where they show quicksort gives a $2$-approximate sorting in $O(n \log n)$ expected comparisons against the non-adaptive adversary. This approach falls apart against the adaptive adversary, however, as if all values are the same, the adversary can force all pivots to compare less than all elements, forcing the algorithm to do $\Omega(n^2)$ comparisons. Our result shows that it is possible to get a constant approximate sorting in near-optimal number of comparisons even against the adaptive adversary.
Previously, this problem had also been studied in the deterministic case \cite{AFHN15}, where an upper bound of $O(4^k\cdot n^{1 + 1/2^{k-1}})$ and a lower bound of $\Omega(n^{1 + 1/2^{k-1}})$ comparisons were proven for $k$-approximate sorting. 
Taking this result with $k = 4$, we get a lower bound of $\Omega(n^{9/8})$ for $4$-approximate deterministic sorting. Thus, our algorithm shows a distinction between randomized and deterministic algorithms in this problem. To get an $\widetilde{O}(n)$ deterministic algorithm, one could at best provide a $\Omega(\log \log n)$-approximation.

Our algorithm uses the fact that randomized quicksort has good comparison complexity if there are not big groups of close elements, as the adversary cannot force the pivots too far away. Thus, if randomized quicksort does not work, there must be a large cluster of close elements that we can exploit. We then estimate the order of each element using a $O(\log n)$ size sample of items, using the existence of this cluster to guarantee our estimates are accurate. Finally, we use these approximate orders to find a partition of the input items, and recursively solve as in quicksort.

Our algorithm also implies a similar selection algorithm.

\begin{restatable}{corollary}{lb}\label{corollary:RSelect}
    For any $i$, there exists a randomized algorithm that takes $O(n \log n)$ comparisons, uses $O(\log n)$ parallel rounds, and returns a $4$-approximate $i$-selection with probability $> 1 - \frac{1}{n^2}$.
\end{restatable}

Similarly, this result is a $\log$ factor away from optimal. Again, the best prior result is of \cite{AFJ18}, where their analysis also shows that quickselect gives a $2$-approximate selection in $O(n)$ expected comparisons against the non-adaptive adversary. Against the adaptive adversary, this approach fails in an identical way to quicksort. Our result shows that it is possible to get a constant approximate selection in near-optimal number of comparisons even against the adaptive adversary.
This was also studied in the deterministic setting \cite{AFHN15}, where an equivalent $\Omega(n^{9/8})$ lower bound was shown, so we also show a distinction between randomized and deterministic in this case. Similarly, any $\widetilde{O}(n)$ deterministic algorithm could at best return a $\Omega(\log \log n)$-approximation.

Our approach is identical to the sorting algorithm, except we only have to recursively solve on the relevant side of the partition, as in quickselect.

Next, we provide a family of algorithms that explore the tradeoff between number of rounds, number of comparisons, and approximation factor in the sorting case.

\begin{restatable}{theorem}{lb}\label{theorem:SR}
    For any integer $d > 0$, there exists a deterministic algorithm that takes $d$ rounds, uses $n^{1 + O(1/d)} d$ comparisons, and returns a $2d$-approximate sorting.
\end{restatable}

Again, any approximate sorting algorithm must be able to correctly sort any list of numbers, so such an algorithm must take $\Omega(n^{1+1/d})$ comparisons \cite{BT83}. Thus, this algorithm is optimal up to a constant factor of $1/d$ in the exponent. However, this constant factor is large, as it arises from the notoriously bad constant of the AKS sorting network \cite{AKS}. The best prior result is the aforementioned deterministic algorithms from \cite{AFHN15}. Their $k$-approximate algorithm uses $\Omega(n^{1-\frac{1}{2^{k}-1}})$ rounds and $O(4^kn^{1+1/2^{k-1}})$ comparisons. Thus, their algorithm uses $\Omega(n^{2/3})$ rounds at best. We drastically improve this by giving algorithms that can use an arbitrarily small number of rounds, which could not be done by any prior algorithm (except for the trivial $1$ round round robin tournament). However, our comparison bound is worse than that of \cite{AFHN15}, as it is not possible to achieve their comparison complexity even for regular sorting in rounds.

Our algorithm uses a connection between this problem and the problem of sorting networks that use a sorting oracle of arity $k$. We use a result based on the AKS sorting network \cite{AKS} that gives sorting networks with asymptotically optimal depth $O(\log_k n)$. We then show that these networks imply good algorithms for adversarial sorting, by showing that each round can incur at most $2$ additional approximation error.

Since the constant factor in the exponent is large, we also provide an asymptotically worse algorithm (with respect to $d$) with smaller constant that is better for small constant $d$.

\begin{restatable}{theorem}{lb}\label{theorem:SRBad}
    For any integer $d > 0$, there exists a deterministic algorithm that takes $d$ rounds, uses $n^{1 + 2/\sqrt{d}} d$ comparisons, and returns a $2d$-approximate sorting.
\end{restatable}

Our final result is an extension of these algorithms to selection algorithms that guarantee a constant approximation.

\begin{restatable}{theorem}{lb}\label{theorem:SelectR}
    For any integer $d > 1$ and $i$, there exists a randomized algorithm that takes $d + O(\log d)$ rounds, uses $n^{1+O(1/d)} d \log{n}$ comparisons, and returns a $202$-approximate $i$-selection with probability $> 1 - \frac{1}{n^2}$.
\end{restatable}

This result uses the previous sorting result, as well as the maximum selection in rounds result from \cite{GKK20}. Similarly, such an algorithm must take $\Omega(n^{1+1/(d + O(\log d))})$ comparisons, so our algorithm is optimal up to a constant factor of $1/d$ in the exponent. The best prior result is again the deterministic selection algorithms of \cite{AFHN15}, but their algorithms similarly use $\Omega(n^{1-\frac{1}{2^k-1}}) = \Omega(n^{2/3})$ rounds. Thus, our algorithm is again a drastic improvement in terms of round complexity. On top of this, for $d = \omega(1)$, our algorithm uses $n^{1 + o(1)}$ comparisons, which still beats the deterministic lower bound of \cite{AFHN15}, regardless of the number of rounds the deterministic algorithm uses. Thus, we show that randomized algorithms can beat the best deterministic algorithms even when restricted to an arbitrarily small number of rounds (as long as it increases with $n$).

Our algorithm repeatedly approximates the $k$-th element by taking $n^{2/3}\log{n}$ random subsets of size $n^{1/3}$, sorting them with depth $d$, and splitting around position $k/n^{2/3}$. This results in us reducing the problem to that of size $n^{5/6}$, which we can then solve with a constant approximation using one of our sorting algorithms. The depth $d$ sorting does not guarantee a good approximation, so we instead use a gap-preserving property of all approximate sorting algorithms to show that this must give a good approximation if there are few elements close to the $k$-th smallest. If there are many close elements, we can instead sample a large subset and estimate the $k$-th smallest directly, which is likely to give us one of the close elements. We then show a method of combining these two algorithms to show it is possible to always get a good approximation.

The following tables summarize the previous results for the problems of approximate sorting and selection along with our contributions.

\begin{table}[H]
\begin{tabular}{||c|c|c|c|c|c||}
\hline
    Paper & Adversary & Randomized? & Approximation & Query Complexity & Round Complexity \\ \hline
    \cite{AFJ18} & Non-Adaptive & Deterministic & $2$ & $O(n \log n)$ & $O(\log n)$\\ \hline
    \cite{AFHN15} & Adaptive & Deterministic & $k$ & $O(4^k n^{1+1/2^{k-1}})$ & $O(n^{1-1/(2^k-1)})$\\ \hline
    Our Paper & Adaptive & Randomized & $4$ & $O(n \log^2 n)$ & $O(\log n)$\\ \hline
    Our Paper & Adaptive & Deterministic &$2d$ & $n^{1+O(1/d)}d$ & $d$\\ \hline
\end{tabular}
\caption{Sorting}
\label{tab:my_label}
\end{table}

\begin{table}[H]
\begin{tabular}{||c|c|c|c|c|c||}
\hline
    Paper & Adversary & Randomized? & Approximation & Query Complexity & Round Complexity \\ \hline
    \cite{AFJ18} & Non-Adaptive & Deterministic &$2$ & $O(n)$ & $O(\log n)$\\ \hline
    \cite{AFHN15} & Adaptive & Deterministic &$k$ & $O(2^k n^{1+1/2^{k-1}})$ & $O(n^{1-1/(2^k-1)})$\\ \hline
    Our Paper & Adaptive & Randomized & $4$ & $O(n \log n)$ & $O(\log n)$\\ \hline
    Our Paper & Adaptive & Deterministic & $2d$ & $n^{1+O(1/d)}d$ & $d$\\ \hline
    Our Paper & Adaptive & Randomized & $202$ & $n^{1+O(1/d)}d\log n$ & $d+O(\log d)$\\ \hline
\end{tabular}
\caption{Selection}
\label{tab:my_label}
\end{table}
\bigskip

\subsection{Related Work}
\label{sec:related}

Imprecise comparisons were first considered by R\'enyi \cite{renyi} and Ulam \cite{ulam} in the setting of binary search. The model described allows for a bounded number of incorrect comparisons. An optimal algorithm for this problem was given by Rivest, Meyer, Kleitman, Winklmann, and Spencer \cite{rivest} that uses $O(\log n)$ comparisons. This problem was considered in the parallel setting by Negro, Parlati, and Ritrovato \cite{negro} where they give optimal algorithms for a fixed number of rounds and errors. 

Binary search has also been considered in the setting where comparisons are incorrect with some probability $p < \frac{1}{2}$ \cite{Hor63, BZ74, 41}. Pelc \cite{pelc} gave an algorithm that uses $O(\log n)$ comparisons and gives the correct answer with probability $1 - \eps$ if $p < \frac{1}{3}$. For $\frac{1}{3} \leq p < \frac{1}{2}$, he gave an algorithm that uses $O(\log^2 n)$ comparisons. A later result from Borgstrom and Kosaraju \cite{borgstrom} implies an optimal $O(\log n)$ algorithm for all $p < \frac{1}{2}$.

Sorting with imprecise comparisons was first considered by Lakshmanan, Ravikuman, and Ganesan \cite{60} in the model where the number of incorrect comparisons is bounded by a function $e(n)$. They gave a lower bound of $\Omega(n\log n + en)$ comparisons and an upper bound of $O(n\log n + en + e^2)$ comparisons. The upper bound was later improved to match the lower bound by Bagchi \cite{13} and Long \cite{64}.

Sorting with comparisons that are incorrect with probability $p < \frac{1}{2}$ was considered by Feige, Peleg, Raghavan, and Upfal \cite{41} where they gave an algorithm that uses $O(n\log(n/\eps))$ queries and gives the correct answer with probability $1 - \eps$.

Another common model is that of sorting networks with faulty comparisons. In the model where $e$ gates may be faulty (i.e. do nothing), Yao and Yao \cite{120} gave an algorithm that uses $O(n \log n + en)$ gates. This model has been further studied in the cases where faulty gates may arbitrarily permute their inputs \cite{10, 63} and where gates are faulty with some probability \cite{100}.

Recently, a comparison model has been considered where some comparisons are not allowed to be made at all. This model was introduced by Huang, Kannan, and Khanna \cite{REE1} where they give a randomized algorithm that uses $\widetilde{O}(n^{3/2})$ comparisons with high probability provided the input is sortable. They also give an algorithm that uses $\widetilde{O}(\min(n/p^2, n^{3/2}\sqrt{p}))$ comparisons if the graph of forbidden comparisons is random with edge probability $1-p$. This was recently improved by Kuszmaul and Narayanan \cite{REE2} who give corresponding algorithms using $\widetilde{O}(\sqrt{nm})$ and $O(n \log (np))$ comparisons respectively.

None of these results apply to our comparison model, as the incorrect comparisons are either bounded or random. In our model, however, there can be any number of incorrect comparisons, which can be chosen adversarially. 
There is a notion of 'closeness' of elements that allows comparisons to be incorrect, for which there does not exist an analogue in other models. We note that if we were to 'disallow' all comparisons between items that differ by at most $1$, using the aforementioned algorithms would give a good approximation. However, this would require additional knowledge of which pairs of elements are sufficiently close, which the algorithm does not have in our model.

Solving comparison-based problems in rounds has also been widely considered \cite{SR5, SR6, SR3, Kruskal83, Leighton84, BollobasH85, VISHKIN, AzarV87, AlonA88a, SR2, BollobasB90, AzarP90, SR7, 41, BravermanMW16, topk, CohenMM20}. For sorting in $k$ rounds, Bollob\'as \cite{SR2} showed that $O(n^{1+1/k}\frac{(\log n)^{2-2/k}}{(\log \log n)^{1-1/k}})$ comparisons are sufficient, while Alon and Azar \cite{SR1} showed that $\Omega(n^{1+1/k} (\log n)^{1/k})$ comparisons are necessary. For merging two sorted arrays in $k$ rounds, Haggkvist and Hell \cite{SR3} showed that $\Theta(n^{1 + 1/(2^k - 1)})$ comparisons is necessary and sufficient. For selecting an element of arbitrary order in $k$ rounds, Pippenger \cite{SR4} showed that $O(n^{1+1/(2^k-1)}(\log n)^{2-2/(2^k-1)})$ comparisons are sufficient, while Alon, Azar, and Vishkin \cite{VISHKIN} showed that $\Omega(n^{1+1/(2^k-1)} (\log n)^{2/(2^k-1)})$ comparisons are necessary for a deterministic algorithm. Bollob\'as and Brightwell showed that, using a randomized algorithm, $O(n)$ comparisons is sufficient to select in $k \geq 4$ rounds with high probability. For sorted top-$m$ in $k$ rounds, Braverman, Mao, and Peres \cite{topk} recently showed that $\widetilde{\Theta}(n^{2/k}m^{(k-1)/k} + n)$ comparisons are both necessary and sufficient. Similar consideration of round complexity has also been done in the setting of best arm selection with multi-armed bandits \cite{ban1, ban2}.

These algorithms do not easily generalize to our setting, as they often heavily rely on the existence of a ``correct'' sorted order that is adhered to by the comparisons, as to derive information from the transitive closure of the comparison graph. We also note that the faulty comparison models considered in \cite{topk, CohenMM20} suffer the same drawbacks as the previously described comparison models when extended to our model.

\section{Techniques}

Throughout, we use the fact that the naive round robin tournament algorithm that does all comparisons (denoted \texttt{Tournament}) guarantees a $2$-approximate sorting \cite{AFJ18}. We also use the fact that any approximate sorting algorithm must preserve gaps in the input set of size at least $1$, as one could move the values on each side of the gap arbitrarily far apart without affecting any comparisons. We say some item $x^*$ is a $k$-left-approximation of another item $x_i$ if $x^* \geq x_i - k$. Similarly, $x^*$ is a $k$-right-approximation of $x_i$ if $x^* \leq x_i + k$. Intuitively, $x^*$ is a left-approximation if it not ``too far left'' of $x_i$, and similarly for a right-approximation. If an element is both a $k$-left-approximation and a $k$-right-approximation of $x_i$ it must be a $k$-approximate $i$-selection.

\subsection{Randomized Sorting}

The main issue with sorting in the adversarial comparison setting is balancing worst-case approximation factor with worst-case number of comparisons. Standard sorting algorithms either always guarantee a good approximation, but can be forced to do many comparisons (i.e. quicksort), or always guarantee few comparisons, but can be forced to be give a bad approximation (i.e. mergesort).  With deterministic algorithms, it was shown in \cite{AFHN15} that it is not possible to get the best of both worlds, where they gave a tradeoff between approximation factor and comparisons. Our algorithm shows that in the randomized case, however, it is possible to be near-optimal in both aspects.

In the style of quicksort, our algorithm aims to partition the input set $X$ into two sets $Y$ and $\overline{Y}$ such that $\overline{Y} \geq_4 Y$. If we can guarantee this in every recursive call, we must return a $4$-approximate sorting, as no pair differing by more than $4$ can ever be put in the wrong order \cite{AFJ18}. To ensure a good bound on the number of comparisons, we also require $|Y|,|\overline{Y}| \geq n/8$.

The first phase of our algorithm attempts to partition using random pivots $O(\log n)$ times. Let $x_L$ and $x_R$ be the $n/8$-th smallest and $n/8$-th largest elements of $X$ respectively. For a fixed pivot $x_p$, if $x_p > x_L + 1$, at least $n/8$ items must compare less than $x_p$. Thus, if $|X \cap [x_L, x_L + 1]| \leq n/4$, by a Chernoff bound, less than half of the pivots we try will have less than $n/8$ items compare less with high probability. Similarly, if $|X \cap [x_R - 1, x_R]| \leq n/4$, less than half of the pivots we try will have less than $n/8$ items compare greater with high probability. If both of these inequalities are satisfied, it follows that we will find a ``good'' partition from one of our pivots with high probability. Otherwise, without loss of generality we assume more than half of the pivots had less than $n/8$ compare less. In this case, we have $|X \cap [x_L, x_L + 1]| > n/4$ with high probability. We will exploit this property in the other phases of the algorithm.

The second phase of our algorithm estimates the order of each element $x_i$ by comparing it to a small subset of $X$. Then, we create a partition by taking $Y$ to be the elements with estimated order less than $n/4$, and $\overline{Y}$ the elements with estimated order at least $n/4$. For some fixed item $x_i < x_L - 1$, there must be $7n/8$ items that compare greater than it, so by a Chernoff bound, $x_i$ ends up in $Y$ with high probability. Similarly, if $x_i > x_L + 2$, there must be $3n/8$ items that compare less than it (since $|X \cap [x_L, x_L + 1]| > n/4$), so by a Chernoff bound it ends up in $\overline{Y}$ with high probability. Thus, by a union bound, $\overline{Y} \geq_3 Y$ with high probability. We note that if we did not have such high density in $[x_L, x_L + 1]$, the elements of order around $n/4$ (which could be arbitrarily far apart value-wise) would be impossible to differentiate with a small sample.

The final phase of our algorithm ensures $|Y|,|\overline{Y}| \geq n/8$. Without loss of generality we assume $|Y| \leq |\overline{Y}|$. If $|Y| \geq n/8$, nothing needs to be done as both sets are sufficiently large. Otherwise, we repeatedly sample $m=O(\log n)$ elements of $\overline{Y}$, sort them, and move the $m/8$ smallest elements to $Y$ until $|Y| \geq n/8$.  Since $|Y| < n/8$ and $|X \cap [x_L, x_L + 1]| > n/4$ before any iteration, there must be at least $n/4$ items $\leq x_L$ that are not in $Y$. Thus, by a Chernoff bound, the $m/8$ smallest elements are all $\leq x_L$ with high probability. Since \texttt{Tournament} incurs error at most $2$, it follows that $|Y| \geq_4 Y$ at the end with high probability. Note that by choosing a random permutation to guarantee disjoint subsets, we can do this sampling in parallel.

\subsection{Sorting in Rounds}

The main issue with sorting in rounds is guaranteeing good comparison complexity. Many of the state of the art algorithms for sorting in rounds heavily rely on the existence of a ``correct'' sorting order to guarantee a low number of comparisons. 

We use low-depth sorting networks of arity $m$ using \texttt{Tournament} to implement the sorting oracle. Since each call to \texttt{Tournament} incurs error at most $2$, each round incurs error at most $2$, so we can show the total error is bounded by $2$ times the number of rounds.

\subsection{Selection in Rounds}

Our algorithm improves the approximation factor from the previous sections algorithm with a small $O(\log d)$ additional round overhead. We use a similar approach to the maximum selection algorithm given in \cite{GKK20}, where we first describe an algorithm that gives a good approximation if there are few elements around the actual answer, then describe an algorithm that gives a good approximation if there are many elements around the actual answer, then show a way to combine them to guarantee a good approximation always. Since we are looking for an element in the middle of the sorted order, however, there is some additional complexity with considering close elements on each side of the desired element.

Let $x_i$ be the actual $i$-th smallest element of $X$. The first part of our algorithm guarantees a good left-side approximation if $|X \cap [x_i - 1, x_i]| \leq \frac{1}{10}n^{2/3}$. Similarly, it guarantees a good right-side approximation if $|X \cap [x_i, x_i + 1]| \leq \frac{1}{10}n^{2/3}$. We aim to partition $X$ into three sets $Z, Y, \Gamma$ such that elements of $Z$ are ``less than'' $x_i$, elements of $\Gamma$ are ``greater than'' $x_i$, and $Y$ is the set of ``candidate'' elements to be $x_i$. We sample $cn^{2/3}\log{n}$ subsets of $X$ of size $n^{1/3}$, each time sorting using the depth $d$ algorithm from the previous subsection. We then take the elements within $n^{1/6}$ of the $k/n^{2/3}$-th element of each subset and add them to $Y$ (the set of candidates). We say that elements in positions to the left of $k/n^{2/3} - n^{1/6}$ are on the left side, and the rest of the elements are on the right side. After sampling all subsets, elements which are not in the set of candidates are partitioned into $Z$ and $\Gamma$ based on how frequently they are on the left side. Assume $|X \cap [x_i - 1, x_i]| \leq \frac{1}{10}n^{2/3}$, the other case is symmetric. In this case, roughly $90\%$ of the sampled subsets will contain no elements in $[x_i - 1, x_i]$. For each of these subsets, since our sorting algorithm must be \emph{gap-preserving}, the sorting must be correct with respect to $[x_i - 1, x_i]$. It thus follows by a tail bound of the Hypergeometric distribution and a union bound that for all $x_j < x_i - 1$, $x_j$ will end up in $Z \cup Y$ with high probability. Similarly, for $x_j > x_i$, $x_j$ will end up in $Y \cup \Gamma$ with high probability. Finally, if $|Z| \geq k$, we return the maximum element of $Z$ computed with a depth $O(\log d)$ maximum finding algorithm from \cite{GKK20}. 
If $|Z| + |Y| < k$, we return the minimum element of $\Gamma$ similarly. Otherwise, we return the $(i-|Z|)$-th element of $Y$ as determined by a constant depth, constant approximate sorting algorithm from the previous section (since $|Y| = O(n^{5/6})$). If $|Z| \geq i$, there must be some element of $Z$ that is $\geq x_i$, so since the maximum finding algorithm returns a constant approximation, we return a constant left-approximation. If $|Z| + |Y| < i$, we return some element of $\Gamma$, and all elements of $\Gamma$ are $\geq x_i - 1$.  Otherwise, there can be at most $i-|Z|-1$ elements of $Y$ that are $<x_i$, so the actual $(k-|Z|)$-th element is $\geq x_i$, so since we use a constant approximate sorting, the element we return is a constant left-approximation as desired.

The second part of our algorithm guarantees a good left side approximation given $|X \cap [x_i - 1, x_i]| > \frac{1}{10}n^{2/3}$ (and there is a symmetric algorithm that guarantees a good right side approximation). The idea is simple: sample $O(n^{5/6})$ elements of $x$, sort them with a constant approximation algorithm in constant rounds, and then take the $(k/n^{1/6} - n^{5/12})$-th element. By another Hypergeometric tail bound, it follows that we always get a good right-approximation, and also get a good left-approximation if $|X \cap [x_i - 1, x_i]| > \frac{1}{10}n^{2/3}$ as desired.

Finally, we describe how to combine these algorithms. First, we run the sparse algorithm and get the result $x^*$. Then, we count the number of elements that compare less than the result. If this value is $\leq i-1$, we must have $x^* \leq x_i + 1$. We then call the left-side dense algorithm and return the greater of the two results (according to the comparator). Since one of the two algorithms must return a constant left-approximation, the latter algorithm always returns a constant right approximation, and we already know that $x^*$ is a constant right approximation, it follows that in the end we return a constant approximation. The case where the number of elements that compare less is $\geq i$ is handled symmetrically.


\section{Preliminaries}
\label{sec:pre}

In this section, we give some basic definitions and results in the adversarial comparison setting which will serve as the basis for many of our algorithms. 

\begin{definition}
    Element $x_j$ in the set $X = \{x_1, \ldots, x_n\}$ is of $k$-order $i$ if there exists a partition $S_1, S_2$ of $X \backslash \{x_j\}$ such that $|S_1| = i-1$, and $S_2 \cup \{x_j\} \geq_k S_1 \cup \{x_j\}$.
\end{definition}

This the notion of approximate selection that was originally introduced in \cite{AFHN15}. We show that this is equivalent to the intuitive notion we previously described, and then show that it is equivalent to something that is easier to work with.

\begin{lemma}
    An item $x_j$ in $X$ is of $k$-order $i$ if and only if $x_j$ is a $k$-approximate $i$-selection.
\end{lemma}
\begin{proof}
    Assume $S_1, S_2$ exist as in the definition of $x_j$ being $k$-order $i$. Then, since $S_2 \cup \{x_j\} \geq_k S_1 \cup \{x_j\}$, the concatenation of the sorted order of $S_1$, $x_j$, and the sorted order of $S_2$ in that order is a $k$-approximate sorting by definition. Thus, $x_j$ is a $k$-approximate $i$-selection as desired.

    Similarly, assume there exists a $k$-approximate sorting $Y$ of $X$ where $y_i = x_j$. Then, taking $S_1$ to be the first $i-1$ elements of $Y$, and $S_2$ to be the final $n-i$ elements, it follows by definition that $S_2 \cup \{x_j\} \geq_k S_1 \cup \{x_j\}$. Thus, $x_j$ is of $k$-order $i$ as desired.
\end{proof}

\begin{restatable}{lemma}{lb}\label{lemma:approxselect}
    An item $x_j$ in $X$ is a $k$-approximate $i$-selection if and only if $|x_j - x_i| \leq k$ where $x_i$ is the actual $i$-th smallest element of $X$.
\end{restatable}

\begin{proof}
    Consider some $k$-approximate sorting $Y$ of $X$ where $y_i = x_j$. Without loss of generality, assume $x_j \leq x_i$. Since $x_i$ is the $i^\text{th}$ smallest element of $X$, there must be at least $n-i+1$ elements of $X$ that are $\geq x_i$. Thus, there must exist some element $x_\ell \geq x_i$ that is to the left of $x_j$ in $Y$ since there are only $n-i$ places to the right that they can be. It follows that $x_j \geq_k x_\ell$, which implies $x_j \geq_k x_i$ since $x_\ell \geq x_i$. Thus, $x_i \geq x_j \geq x_i - k$, so $|x_j-x_i| \leq k$.

    Consider some element $x_j$ in $X$ such that $|x_j - x_i| \leq k$. Let $Y$ be the sorted order of $X$, but with $x_j$ and $x_i$ swapped. All pairs that do not contain $x_i$ or $x_j$ must still be in the right order. Pairs $(x_i, x_\ell)$ that are in the wrong order must have $x_\ell$ between $x_i$ and $x_j$ (or equal to $x_j$), so $x_\ell$ and $x_i$ must differ by at most $k$. An identical argument applies to pairs $(x_j, x_\ell)$, so it follows that $Y$ is a $k$-approximate sorting as desired.
\end{proof}

\begin{restatable}{corollary}{lb}\label{corollary:sortdiff}
    If $Y$ is a $k$-approximate sorting of $X$, and $S$ is the actual sorting of $X$, $|y_i - s_i| \leq k$ for all $i$.
\end{restatable}

We define the notion of a \emph{gap-preserving} algorithm, where elements must be correctly sorted with respect to a gap of size $1$. This will be useful in proving the correctness of our later algorithms.

\begin{definition}
    We say a sorting algorithm is \emph{gap-preserving} if, given there exists a gap of length $1$ in the input, the sorting algorithm returns all elements before the gap before all elements after the gap. Formally, given input $X$ such that there exists a gap $[y,y+1)$ where $X \cap [y,y+1) = \emptyset$, the sorting algorithm must return all elements of $X$ less than $y$ before all elements of $X$ greater than $y$. 
\end{definition}

\begin{restatable}{lemma}{lb}\label{lemma:jumppreserving}
    All approximate sorting algorithms are \emph{gap-preserving}.
\end{restatable}
\begin{proof}
    We can shift the elements on one side of the gap arbitrarily far away without affecting any comparison results. Thus, if there exists an input for which a $\tau(n)$-approximate algorithm is not gap-preserving, then we can make the approximation factor larger than $\tau(n)$ by shifting one side by more than that, a contradiction.
\end{proof}

Recall that the \texttt{Tournament} algorithm for sorting a set $X$ of items, as originally defined in \cite{AFHN15}, does all pairwise comparisons, and orders the items by the number of ``wins'' they have (i.e. the number of elements that compare less).

\begin{restatable}{lemma}{lb}\label{lemma:tournament}
    \texttt{Tournament} is a $2$-approximate sorting algorithm.
\end{restatable}

\begin{proof}
    If $x_i > x_j + 2$, $x_i$ must compare greater than all elements $\leq x_j + 1$ including $x_j$. However, $x_j$ can at most  compare greater than all elements $\leq x_j + 1$ excluding itself, so it must come before $x_i$ as desired.
\end{proof}

We show that partitioning as in quicksort guarantees a good approximation factor, which will be the basis of our randomized sorting algorithm. This was originally shown in \cite{AFJ18}.

\begin{restatable}{lemma}{lb}\label{lemma:pivot}
    Let $x_i$ be some item in a set $X$. Let $S = \{x_j \mid x_j <_c x_i\}$ and $T = X \backslash S$. We must have $T \geq_2 S$.
\end{restatable}

\begin{proof}
    All elements of $S$ must be $\leq x_i + 1$, and all elements of $T$ must be $\geq x_i - 1$. Thus, for $x_j \in S, x_k \in T$, $x_j - x_k \leq x_i + 1 - (x_j - 1) = 2$ as desired.
\end{proof}

\begin{restatable}{lemma}{lb}\label{lemma:quicksort}
    If a sorting algorithm repeatedly partitions the input set $X$ into two sets $S,T$ such that $T \geq_k S$, recursively sorts $S$ and $T$ and then concatenates them, it is guaranteed to result in a $k$-approximate sorting.
\end{restatable}
\begin{proof}
    Assume for the sake of contradiction that there exist $x_i,x_j$ for $i > j$ in the final order such that $x_i \not \geq_k x_j$. We must have put $x_i$ in $T$ and $x_j$ in $S$ in some recursive call, but this contradicts $T \geq_k S$, as desired.  
\end{proof}

Throughout, when referring to a sorted order, we assume a fixed sorted order with ties broken arbitrarily. Unless otherwise stated, all logarithms are base $e$. We often ignore rounding errors that vanish for large $n$.

\section{A Randomized Sorting (and Selection) Algorithm}
\label{sec:lb}

In this section we prove Theorem \ref{thm:RSort} by describing an algorithm \texttt{RSort}. This algorithm is similar to quicksort in the sense that we aim to partition the original set of items $X$ into two sets $S,T$, and then recursively sort $S$ and $T$ and concatenate them. Recall by Lemma \ref{lemma:quicksort} that it is sufficient to have $T \geq_4 S$ in every call.
Thus, our algorithm aims to find a partition $S,T$ of $X$ such that $T \geq_4 S$. To ensure the recursion depth is $O(\log |X|)$, we also aim to have $|S|, |T| \geq \frac{|X|}{8}$. Our algorithm consists of three phases, which we will analyze independently. Throughout the algorithm, we let $n$ be the size of the current set $X$ the function is being called on, and we let $N$ be the size of the initial set $X$ that \texttt{RSort} was called on. This distinction is important, as we want our probability guarantees to be with respect to the size of the original caller.

\subsection{The Pivot Phase}

\begin{algorithm}
    \caption{Pivot Phase}
    \label{alg:pivot}
    \begin{algorithmic}[1]
        \State $R \gets 0$
        \State $L \gets 0$
        \Loop \ $8c_1\log N$ times
        \State pick a pivot $x_p$ at random
        \State $Y \gets \{x \in X : x <_c x_p\}$
        \State $\overline{Y} \gets X \backslash Y$
        \If {$\min(|Y|,|\overline{Y}|) \geq \frac{n}{8}$}
        \State \Return $(Y,\overline{Y})$
        \ElsIf {$|Y| < \frac{n}{8}$}
        \State $L \gets L + 1$
        \Else 
        \State $R \gets R + 1$
        \EndIf
        \EndLoop
    \end{algorithmic}
\end{algorithm}

In this phase, we aim to use a pivot as in quicksort to find the desired partition, in which case we return early. If we do not find such a pivot, the input set has additional structure with high probability, which we will use in the rest of the algorithm. 

$8c_1\log N$ elements $x_p$ are randomly chosen and used as pivots. This splits $X$ into two sets $Y$ and $\overline{Y}$ such that $\overline{Y} \geq_2 Y$ by Lemma~\ref{lemma:pivot}. If both of these sets are sufficiently large, then we have found our desired partition, and we return early. Otherwise, if $|Y| < \frac{n}{8}$, we say $x_p$ \emph{goes left} and if $|\overline{Y}| < \frac{n}{8}$, we say $x_p$ \emph{goes right}. Let $x_L$ be the $n/8$-th smallest element of $X$ and $x_R$ the $n/8$-th largest. If $x_p > x_L + 1$, then $x_p$ cannot go left, and symmetrically, if $x_p < x_R - 1$, then $x_p$ cannot go right. Let $X_L = \{ x_p \in X : x_p \leq x_L + 1 \}$ and $X_R = \{ x_p \in X : x_p \geq x_R - 1\}$. The elements in $X \backslash (X_L \cup X_R)$ are thus guaranteed to neither go left or go right. Intuitively, if this set is sufficiently big, we expect to find such a pivot. Otherwise, either $X_L$ or $X_R$ must be large. The variables $L$ and $R$ in the code count how many pivots go left and right respectively. Again intuitively, we expect $L > R$ if $X_R$ is small and vice versa. These intuitive statements are captured in the following lemmas:

\begin{lemma}
If $|X_L| < \frac{3n}{8}$, for any constant $r>0$, we can choose $c_1$ sufficiently large such that $\pr{L \geq 4c_1\log N \textup{ after pivot phase}} < \frac{1}{N^r}.$
\end{lemma}
\begin{proof}
Let $A_i$ be a random variable that takes value $1$ if the $i^{\text{th}}$ pivot $x_p$ is in $X_L$, and $0$ otherwise. Note that if $A_i$ is $0$, we cannot increment $L$ in the $i^{\text{th}}$ iteration, so we have $L \leq A = \sum_i A_i$. Let $\mu = \mathbb{E}[A] = 8c_1\log N\frac{|X_L|}{n} \geq c_1\log N$. By a Chernoff bound, we have:
\begin{align*}
\pr{L \geq 4c_1\log N \textup{ at line }20} &\leq \pr{A \geq 4c_1\log N}\\
&= \pr{A \geq (1 + \delta)\mu}
\shortintertext{Where $\delta = \frac{n}{2|X_L|}-1 > \frac{1}{3}$}
&\leq e^{-\frac{\delta^2\mu}{2+\delta}}\\
&< e^{-\frac{\mu}{21}}\\
&\leq e^{-\log N\frac{c_1}{21}}\\
&= N^{-\frac{c_1}{21}}
\shortintertext{Thus, choosing $c_1\geq 21r$, we get}
\pr{L \geq  4c_1\log N \textup{ at line }20} &< \frac{1}{N^r}.
\end{align*}
\end{proof}
\begin{corollary}
If $|X_R| < \frac{3n}{8}$, for any constant $r>0$, we can choose $c_1$ sufficiently large such that $\pr{R \geq 4c_1\log N \textup{ after pivot phase}} < \frac{1}{N^r}.$
\end{corollary}
\begin{proof}
    Symmetric.
\end{proof}

Throughout the rest of the analysis, we will assume $L \geq R$. The other case is handled symmetrically.

\subsection{The Sample Phase}

\begin{algorithm}
    \caption{Sample Phase}
    \label{alg:pivot}
    \begin{algorithmic}[1]
        \State $Y \gets \emptyset$
    \For {$x_i \in X$}
        \State $C \gets 0$
        \Loop \ $8c_2\log N$ times
            \State Choose $z \in X$ at random
            \If {$z <_c x_i$}
                \State $C \gets C + 1$
            \EndIf
        \EndLoop
        \If {$C < 2c_2\log N$}
            \State $Y \gets Y \cup \{x_i\}$
        \EndIf
    \EndFor
    \end{algorithmic}
\end{algorithm}

In this phase, for each element $x_i$ we estimate its position in the sorted array by comparing it to a small subset of $X$. All elements with estimated position less than $\frac{n}{4}$ are put in set $Y$ and the remaining elements are put in $\overline{Y}$. Since $|X_L| \geq \frac{3n}{8}$, all elements with positions between $\frac{n}{8}$ and $\frac{3n}{8}$ are in $[x_L, x_L + 1]$. Thus, since all elements $< x_L - 1$ compare less than all of these elements, we intuitively expect them to have estimated position less than $\frac{n}{4}$ even on a small subset. Similarly, since all elements $> x_L + 2$ compare greater than all of those elements, we intuitively expect them to have estimated position greater than $\frac{n}{4}$. These statements are captured in the following lemmas:

\begin{lemma}
If $|X_L| \geq \frac{3n}{8}$, for any constant $r>0$ we can choose $c_2$ sufficiently large such that \[\pr{\exists x_i < x_L - 1 : C \geq 2c_2\log N} < \frac{1}{N^r}.\]
\end{lemma}
\begin{proof}
Let $U$ be the set of the smallest $n/8$ elements of $X$. Consider some iteration of the loop on line $25$ where $x_i < x_L - 1$. Let $A_i$ be a random variable that takes value $1$ if the $i^{\text{th}}$ random element is $\in U$ and $0$ otherwise. Note that if $A_i$ is $0$, we cannot increment $C$ in the $i^{\text{th}}$ iteration. Thus, $C \leq A = \sum_i A_i$. Let $\mu = \mathbb{E}[A] = 8c_2\log N\frac{|U|}{n} = c_2 \log N$, we have:
\begin{align*}
\pr{C \geq 2c_2\log N} &\leq \pr{A \geq 2c_2\log N}\\
&= \pr{A \geq (1+\delta)\mu}
\shortintertext{Where $\delta = 1$}
&\leq e^{-\frac{\delta^2 \mu}{2 + \delta}}\\
&= e^{-\frac{\mu}{3}}\\
&= e^{-\log N \frac{c_2}{3}}\\
&= N^{-\frac{c_2}{3}}
\shortintertext{Thus, choosing $c_2 \geq 3(r+1)$, we get}
\pr{C \geq 2c_2\log N} &\leq \frac{1}{N^{r+1}}
\shortintertext{By a union bound:}
\pr{\exists x_i < x_L - 1: C \geq 2c_2\log N} &\leq \#\{x_i < x_L - 1\}\frac{1}{N^{r+1}}\\
&< \frac{1}{N^r}.
\end{align*}
\end{proof}
\begin{lemma}
If $|X_L| \geq \frac{3n}{8}$, for any constant $r>0$ we can choose $c_2$ sufficiently large such that \[\pr{\exists x_i > x_L + 2 : C < 2c_2\log N} < \frac{1}{N^r}.\]
\end{lemma}
\begin{proof}
Similar to the previous Lemma, using the fact that $|X_L| \geq \frac{3n}{8} \implies \#\{x_i \leq x_L + 1\} \geq \frac{3n}{8}$.
\end{proof}
\begin{corollary}
If $|X_L| \geq \frac{3n}{8}$, for any constant $r>0$ we can choose $c_2$ sufficiently large such that after the sample phase, $\pr{\max(Y) > x_L + 2} < \frac{1}{N^r}$.
\end{corollary}
\begin{proof}
If $\max(Y) > x_L + 2$ then we must have had $C < 2c_2\log N$ for some $x_i > x_L + 2$, which happens with probability $< \frac{1}{N^r}$ by the previous Lemma.
\end{proof}
\begin{corollary}
If $|X_L| \geq \frac{3n}{8}$, for any constant $r>0$ we can choose $c_2$ sufficiently large such that after the sample phase, $\pr{\min(X\backslash Y) < x_L - 1} < \frac{1}{N^r}$.
\end{corollary}

\subsection{The Shifting Phase}

\begin{algorithm}[H]
    \caption{Shifting Phase}
    \label{alg:pivot}
    \begin{algorithmic}[1]
    \State Let $P$ be a random permutation of $X \backslash Y$
    \State $i \gets 0$
    \State $B \gets 4c_3 \log N$
    \While {$|Y| < \frac{n}{8}$}
		\State $Z \gets P[i..i+7B)$
		\State Tournament$(Z)$
        \State $Y \gets Y \cup Z[0..B)$
        \State $i \gets i + 7B$
    \EndWhile
    \State Let $P$ be a random permutation of $Y$
    \State $i \gets 0$
    \While {$|Y| > \frac{7n}{8}$}
        \State $Z \gets P[i..i+7B)$
		\State $Z \gets $ Tournament$(Z)$
        \State $Y \gets Y \backslash Z[6B..7B)$
        \State $i \gets i + 7B$
    \EndWhile
    \State Let $\overline{Y} = X\backslash Y$
    \State \Return $(Y, \overline{Y})$
    \end{algorithmic}
\end{algorithm}

In this phase, if either $Y$ or $\overline{Y}$ is too big, we move some elements to the other set to ensure they both have size $\geq n/8$. Since the two cases are symmetric, without loss of generality, we assume $|Y| \leq |\overline{Y}|$. We partition $\overline{Y}$ into small subsets, and move the minimum $1/8$-th of each subset into $Y$ until $|Y| \geq n/8$. Since at least $3n/8$ elements of $X$ are $\leq x_L + 1$, at least $1/4$-th of the elements in $\overline{Y}$ are $\leq x_L + 1$, so even for small subsets we expect the smallest $1/8$-th to be all $\leq x_L + 1$.  Thus, since \texttt{Tournament} returns a $2$-approximate sorting by Lemma \ref{lemma:tournament}, we expect the elements we add to $Y$ to be $\leq x_L + 3$. These intuitive statements are captured in the following lemmas:

\begin{lemma}
If $|X_L| \geq \frac{3n}{8}$ and $\max(Y) \leq x_L + 2$ after the sample phase, for any constant $r>0$ we can choose $c_3$ sufficiently large such that after the shifting phase, $\pr{\max(Y) > x_L + 3} < \frac{1}{N^r}$
\end{lemma}
\begin{proof}
Consider some iteration of the loop on line $40$. Recall that \texttt{Tournament} returns a 2-approximate sorting. Thus, if in each iteration of the loop, $Z$ has at least $B$ elements $\leq x_L + 1$, then we are guaranteed to only add elements $\leq x_L + 3$ to $Y$. Since $|X_L| \geq \frac{3n}{8}$ and $|Y| < \frac{n}{8}$, there must be at least $\frac{n}{4}$ elements $x \in X \backslash Y$ such that $x \leq x_L + 1$. Let $U$ be the set of the $\frac{n}{4}$ smallest elements of $X \backslash Y$, breaking ties arbitrarily. 
Let $A_i$ be a random variable that takes value $1$ if $Z_i \not \in U$ (before sorting), and $0$ otherwise. Note that for any subset $S$ of $\{A_i\}$, \[\pr{\bigwedge_{i \in S} A_i} = \pr{A_{S_0}}\pr{A_{S_1} | A_{S_0}}\ldots \pr{A_{S_{|S|}} | A_{S_0},\ldots,A_{S_{|S|-1}}} \leq \frac{3}{4} \frac{\frac{3|X \setminus Y|}{4}-1}{|X \setminus Y|-1}\ldots \frac{\frac{3|X \setminus Y|}{4}-|S|}{|X \setminus Y|-|S|} \leq \left (\frac{3}{4}\right )^{|S|}.\] Let $C = \#\{x_i \in Z | x_i > x_L + 1\}$. Clearly, $x_i > x_L + 1 \implies x_i \not \in U$, so $C \leq A = \sum_i A_i$. We have:
\begin{align*}
\pr{C \geq 6B} &\leq \pr{A \geq 6B}\\
&\leq e^{-7B\left (2\left (\frac{6}{7}-\frac{3}{4}\right )^2 \right)}\\
&= e^{-\frac{9B}{56}}\\
&= e^{-\log N \frac{9c_3}{14}}\\
&= N^{-\frac{9c_3}{14}}
\shortintertext{Thus, choosing $c_3 \geq \frac{14(r+1)}{9}$}
&\leq \frac{1}{N^{r+1}}
\shortintertext{By a union bound:}
\pr{C \geq 5B \text{ on some iteration}} &\leq \frac{n}{B}\frac{1}{N^{r+1}}\\
&< \frac{1}{N^r}.
\end{align*}
Here we use the generalized Chernoff bound from Theorem 1.1 of \cite{IK10}.
\end{proof}

Since we initially have $|Y| < |\overline{Y}|$, elements are only added in this case, so if we had $\min(X \setminus Y) \geq x_L - 1$ before this phase, this must still be the case. If this happens, and the conclusion of Lemma 4.7 holds, then after this phase, $\max(Y) \leq x_L + 2$ and $\min(X \setminus Y) \geq x_L - 1$, so $\overline{Y} \geq_4 Y$. Furthermore, after this phase, both $Y$ and $\overline{Y}$ must have size at least $\frac{n}{8}$, as otherwise we would have added elements to the smaller one in this phase. Thus, if these conditions hold, then we get the desired result.

\subsection{Tying it together}

We conclude the probability bounds for the algorithm and describe the comparison and round complexity.

\begin{lemma}
For any constant $r>0$, we can choose $c_1, c_2, c_3$ sufficiently large such that the probability that we split $X$ into sets $Y,\overline{Y}$ such that $\overline{Y} \geq_4 Y$ is $> 1 - \frac{1}{N^r}$.
\end{lemma}
\begin{proof}
    As described in the previous sections, there are $6$ failure points at which something may go wrong and we may end up with $\overline{Y} \not \geq_4 Y$. By a union bound, it follows that the probability that $\overline{Y} \geq_4 Y$ is at least $1 - \frac{6}{N^{r+1}} > 1-\frac{1}{N^r}$ for sufficiently large $c_1, c_2, c_3$ as desired.
\end{proof}

\begin{theorem}
For any constant $r>0$, we can choose $c_1, c_2, c_3$ sufficiently large such that \texttt{RSort} returns a $4$-approximate sorting with probability $>1 - \frac{1}{N^r}$.
\end{theorem}
\begin{proof}
Recall that it is sufficient for every recursive call to satisfy $\overline{Y} \geq_4 Y$. Since we reduce the size of the input by a constant factor in each recursive call, there must be $O(N)$ total recursive calls. Thus, by a union bound, we return a $4$-approximate sorting with probability at least $1 - O(N)\frac{1}{N^{r+2}} > 1 - \frac{1}{N^r}$ as desired.
\end{proof}

\begin{theorem}
\texttt{RSort} uses $O(N \log^2 N)$ comparisons.
\end{theorem}
\begin{proof}
It is clear that any recursive call takes $O(n \log N)$ comparisons. It follows by a well known recurrence that $O(N \log^2 N)$ comparisons are thus required in total.
\end{proof}

\begin{theorem}
\texttt{RSort} uses in $O(\log N)$ rounds.
\end{theorem}
\begin{proof}
The different iterations of each loop in \texttt{RSort} are clearly independent, so we can do them in parallel. Thus each call to \texttt{RSort} takes $O(1)$ rounds. Additionally, each layer of the recursion can also be done in parallel. Since we reduce the size of the input by a constant factor in each call, the recursion depth is $O(\log N)$ and thus the algorithm works in $O(\log N)$ rounds.
\end{proof}

Theorem~\ref{thm:RSort} thus follows from the previous three Theorems.

\medskip
By only recursively solving on the relevant side, this sorting algorithm implies a selection algorithm that returns a $4$-approximation with probability $> 1 - \frac{1}{N^r}$ that uses $O(N \log N)$ comparisons and $O(\log N)$ rounds. Corollary~\ref{corollary:RSelect} thus follows.


\section{A General Sorting Algorithm In Rounds}
\label{sec:non-adaptive}

In this section, we use a connection to sorting networks to give a general sorting algorithm in rounds. We consider sorting networks of arity $k$: rather than being able to compare and swap two elements, we can sort any $k$ elements.

\begin{theorem} 
    \cite{SNLEC} For all $m \geq 2$, there exists an arity $m$ sorting network of depth $O(\log_m n)$.
\end{theorem}

\begin{corollary}
    For any integer $d > 0$, there exists a sorting network of arity $n^{O(1/d)}$ and depth $d$.
\end{corollary}

This result comes from the AKS sorting network construction \cite{AKS}, which has a notoriously big constant factor. Thus, we also consider asymptotically worse (with respect to $d$) networks with smaller constant factors, which are better for small $d$.

\begin{theorem}
    \cite{SN2} For all $m \geq 2$, there exists an arity $m$ sorting network of depth $4\log^2_m n$.
\end{theorem}

\begin{corollary}
    For any integer $d > 0$, there exists a sorting network of arity $n^{2/\sqrt{d}}$ and depth $d$.
\end{corollary}

We connect this result to the adversarial comparison setting by showing that these sorting networks imply approximate sorting algorithms. Since \texttt{Tournament} gives a $2$-approximate sorting, by implementing the sorting oracle with \texttt{Tournament}, we in some sense guarantee that the total approximation error only accumulates by $2$ on each level of the network. Thus, for a depth $d$ network, we get a $2d$-approximate algorithm. 

\begin{lemma}
    Let $a$ and $b$ be arrays of length $n$. If $|a_i - b_i| \leq k$ for all $i$, then $|\text{sorted}(a)[i] - \text{sorted}(b)[i]| \leq k$ for all $i$.
\end{lemma}
\begin{proof}
    We proceed by induction over $n$. When $n = 1$, the result is trivial. Otherwise, let $i = \argmin(a)$, $j = \argmin(b)$. Without loss of generality, assume $a[i] \leq b[j]$. If $i = j$, then $|\text{sorted}(a)[0] - \text{sorted}(b)[0]| \leq k$ and the result follows by the induction hypothesis. Otherwise, we claim that $|b[i] - a[j]| \leq k$. If $b[i] \geq a[j]$, then $|b[i] - a[j]| = b[i] - a[j] \leq b[i] - a[i] \leq k$. Otherwise, $|b[i] - a[j]| = a[j] - b[i] \leq a[j] - b[j] \leq k$. Thus, we can swap $a[i]$ and $a[j]$ and the assumption still holds. We thus reduce to the already solved $i = j$ case as desired.
\end{proof}

\begin{lemma}
If there exists a sorting network with arity $k$ and depth $d$, then there exists a $2d$-approximate sorting algorithm in $d$ rounds that takes $O(nkd)$ comparisons.
\end{lemma}
\begin{proof}
Consider directly running the sorting network, using \texttt{Tournament} to sort. Clearly, $O(dn/k)$ groups are sorted, and each takes $O(k^2)$ time, so the total time taken is $O(nkd)$. We claim that after the $r$-th round, the current element at position $i$ differs by the ``correct'' element at position $i$ (the element that would be there if all comparisons were correct) by at most $2r$. We prove this by induction. When $r = 0$, the result is trivial. Otherwise, after $r-1$ rounds, each element must differ by at most $2r-2$ from the ``correct'' element. By the previous lemma, it follows that in each group that is being sorted, the elements of the correct sorting of the current elements differ by the elements of the correct sorting of the correct elements by at most $2r - 2$. Since \texttt{Tournament} gives a $2$-approximate sorting, it follows by Corollary \ref{corollary:sortdiff} that after sorting the elements differ by the ``correct'' elements by at most $2r$ by the triangle inequality as desired.
\end{proof}

Theorem~\ref{theorem:SR} and Theorem~\ref{theorem:SRBad} follow. 
By letting $d$ be an arbitrarily large constant, we can get a constant round, constant approximate algorithm that uses $O(n^{1+\eps})$ comparisons for any $\eps > 0$.
 
\section{A General Selection Algorithm In Rounds}
\label{sec:comparators}
In this section, we extend the sorting algorithms in the previous section to selection algorithms that return a constant approximation regardless of $d$. We first provide an algorithm that gives a good approximation if there are few elements close to the answer. Then, we provide an algorithm that gives a good approximation if there are many elements close to the answer. We then show that it is possible to combine these to always achieve a constant approximation.

\subsection{Sparse Selection}

Let $L_x = \{x_i | x_k - 1 \leq x_i \leq x_k\}$ and $R_x = \{x_i | x_k \leq x_i \leq x_k + 1\}$. This part of the algorithm returns a $200$-approximation on the left side if $|L_x|$ is sufficiently small, and a $200$-approximation on the right side if $|R_x|$ is sufficiently small. Specifically, if $|L_x| \leq \frac{1}{10}n^{2/3}$, then $x^* \geq_{200} x_i$ where $x^*$ is the returned item. Similarly, if $|R_x| \leq \frac{1}{10}n^{2/3}$, then $x_i \geq_{200} x^*$.

We aim to partition $X$ into three sets: $Z,Y,\Gamma$ where $Z$ is the set of elements definitely to the left of $x_k$, $Y$ is the set of candidate elements to be $x_k$, and $\Gamma$ is the set of elements definitely to the right of $x_k$. We also want $|Y| = O(n^{1-\eps})$, so we can sort $Y$ with a constant approximate algorithm. We sample $cn^{2/3}\log{n}$ subsets of $X$ of size $n^{1/3}$, each time sorting with the $d$ round algorithm from the previous section. We then take the elements of each subset close to the $k/n^{1/3}$-th position and add them to the set of candidates. The elements that are not candidates at the end are partitioned into left and right depending whether they were to the left or the right of the $k/n^{1/3}$-th position more frequently. If $|L_x|$ is sufficiently small, we expect most of the subsets to not contain any elements of $L_x$, and thus since \texttt{Sort} must be \emph{gap-preserving}, the subsets must be roughly correctly sorted around position $k/n^{1/3}-\frac{|L_x|}{2}$. Thus, we expect our candidates to be $\geq x_k - 1$. Similarly, when $|R_x|$ is sufficiently small, we expect our candidates to be $\leq x_k + 1$. This gives us our desired result.

Let $d > 1$ be arbitrary. Let \texttt{Sort} be the $d$ round sorting algorithm from Theorem \ref{theorem:SR} and let \texttt{100-Sort} be the sorting algorithm obtained by taking $d = 100$ in Theorem \ref{theorem:SRBad}. 

\begin{theorem}
    \cite{GKK20} For any $r > 0$, there exists a $\log_2 d$ round maximum/minimum finding algorithm \texttt{GetMax}/\texttt{GetMin} that uses $O(n^{1 + \frac{1}{d-1}}\log d\log n)$ comparisons and returns a $5$-approximate maximum/minimum with probability $> 1 - \frac{1}{n^r}$\footnote{Their algorithm guarantees a $3$-approximation, but only probability $\geq 0.9$. Running it $O(\log n)$ times and taking the tournament maximum/minimum of the results boosts the probability to $>1-\frac{1}{n^r}$, but increases the approximation factor to $5$.}.
\end{theorem}

\begin{algorithm}[H]
\caption{Sparse Selection}\label{alg:cap}
\begin{algorithmic}[1]
\Function{SelectSparse}{$X, k$} 
    \State $Y \gets \emptyset$
    \State $L \gets [0] * n$
    \Loop { $cn^{2/3}\log{n}$ times}
        \State Generate a subset $S$ of $X$ of size $n^{1/3}$
        \State $S \gets$ Sort$(S)$
        \State $T \gets S[k/n^{2/3} - n^{1/6} : k/n^{2/3} + n^{1/6}]$
        \For {$x_i \in S[:k/n^{2/3}-n^{1/6}]$}
            \State $L[i] \gets L[i] + 1$
        \EndFor
        \State $Y \gets Y \cup T$
    \EndLoop
    \State $Z \gets \emptyset$
    \For {$i = 0..n-1$}
        \If {$x_i \not \in Y$ and $L[i] > \frac{c}{2} \log{n}$}
            \State $Z \gets Z \cup \{x_i\}$
        \EndIf
    \EndFor
    \State $\Gamma \gets X \backslash (Y \cup Z)$
    \If {$k \leq |Z|$}
        \State \Return GetMax$(Z)$
    \ElsIf {$k \leq |Z| + |Y|$}
        \State $Y \gets$ 100-Sort$(Y)$
        \State \Return $Y[k-|Z|-1]$
    \Else
        \State \Return GetMin$(\Gamma)$
    \EndIf
\EndFunction
\end{algorithmic}
\end{algorithm}

\begin{lemma}
    If $|L_x| \leq \frac{1}{10} n^{2/3}$, for $r>0$ and $x_i > x_L$ there exists $c$ large enough that $\pr{L[i] > \frac{c}{2}\log n} < \frac{1}{n^r}$.
\end{lemma}
\begin{proof}
    Consider the iterations in which $x_i$ is chosen. By a union bound,
    \[\pr{L_x \cap S \neq \emptyset \mid x_i \in S} \leq |L_x|\frac{n^{1/3} - 1}{n - 1} \leq \frac{1}{10} n^{2/3} \frac{n^{1/3} - 1}{n - 1} \leq \frac{2}{10}\] for $n$ sufficiently large. Conditioning on $x_i$ being $\in S$, the number of elements of $S$ that are $\leq x_k$ (call this $V$) follows a Hypergeometric$(n, k, n^{1/3})$\footnote{If there are multiple items with value $x_k$, the second argument can be larger than $k$, but that can only make the bounds better.} distribution. By a tail bound:
    \begin{align*}
        \pr{V \leq k/n^{2/3} - n^{1/6} \mid x_i \in S} &= \pr{V \leq (k/n-n^{-1/6})n^{1/3}}\\ 
        &\leq e^{-2(n^{-1/6})^2n^{1/3}}\\
        &= e^{-2}.
    \end{align*}
    If $L_x \cap S = \emptyset$ and $V > k/n^{2/3} - n^{1/6}$, since \texttt{Sort} is \emph{gap-preserving} by Lemma \ref{lemma:jumppreserving}, $L[i]$ cannot increase in this iteration. It thus follows by a union bound that $L[i]$ increases with probability at most $\frac{2}{10} + e^{-2} < 0.4$. Thus, $\pr{x_i \in S \text{ and } L[i] \text{ increases}} < \frac{0.4}{n^{2/3}}$. Let $\mu = \mathbb{E}[L[i]] \leq cn^{2/3}\log n \frac{0.4}{n^{2/3}} = 0.4 c \log n$. By a Chernoff bound:
    \begin{align*}
        \pr{L[i] > \frac{c}{2}\log n} &\leq \pr{L[i] > (1 + 1/4)\mu}\\
        &\leq e^{-(1/4)^2 \mu / (2 + 1/4)}\\
        &\leq e^{-0.4c\log n / 36}\\
        \shortintertext{Choosing $c > 90r$:}
        &< \frac{1}{n^r}
    \end{align*}
    as desired.

\end{proof}

\begin{corollary}
    If $|L_x| \leq \frac{1}{10} n^{2/3}$, for any $r > 0$ there exists $c$ large enough that $\pr{Z \cap (x_L, \infty) = \emptyset} > 1 - \frac{1}{n^r}$.
\end{corollary}
\begin{proof}
    This follows by a union bound and the previous lemma.
\end{proof}

\begin{corollary}
    If $|L_x| \leq \frac{1}{10} n^{2/3}$, for any $r > 0$ there exists $c$ large enough that $\pr{\Gamma \cap (-\infty, x_L - 1) = \emptyset} > 1 - \frac{1}{n^r}$.
\end{corollary}
\begin{proof}
    Symmetric.
\end{proof}

Let $x^*$ be the value returned by \texttt{SelectSparse}.

\begin{lemma}
    If $|L_x| \leq \frac{1}{10} n^{2/3}$, for any $r > 0$ there exists $c$ large enough that $\pr{x^* \geq_{200} x_k} > 1 - \frac{1}{n^r}$.
\end{lemma}
\begin{proof}
    We claim it suffices that $Z \cap (x_L, \infty) = \emptyset$ and $\Gamma \cap (-\infty, x_L - 1) = \emptyset$. If $|Z| \geq k$, then there must be an element of $Z$ that is $\geq x_k$. Thus, since \texttt{GetMax} returns a $5$-approximation with sufficiently large probability, $x^* \geq_5 x_k$ with sufficiently large probability. If $|Z| + |Y| < k$, then we return some element of $\Gamma$ which is $\geq_1 x_k$ if $\Gamma \cap (-\infty, x_L - 1) = \emptyset$. Otherwise, if $|Z| < k$ and $|Z| + |Y| \geq k$, there must be at most $k-|Z|-1$ elements of $Y$ that are $< x_k$. Thus, the $(k-|Z|)$-th element of $Y$ is $\geq x_k$. Since \texttt{100-Sort} returns a $200$-approximation, it follows that $x^* \geq_{200} x_k$ with sufficiently large probability as desired.
\end{proof}

\begin{corollary}
    If $|R_x| \leq \frac{1}{10} n^{2/3}$, for any $r > 0$ there exists $c$ large enough that $\pr{x_k \geq_{200} x^*} > 1 - \frac{1}{n^r}$.
\end{corollary}
\begin{proof}
    Symmetric.
\end{proof}

\begin{theorem}
    \texttt{SelectSparse} uses $n^{1+O(1/d)}d\log{n}$ comparisons and $d + \max(100, \log_2 d)$ rounds.

\end{theorem}

\begin{proof}
    All of the comparisons come from \texttt{Sort}, \texttt{100-Sort}, and \texttt{GetMax}/\texttt{GetMin}. The number of comparisons is thus bounded by $cn^{2/3}\log{n}(n^{1/3})^{1+O(1/d)} + n^{1+\frac{1}{d-1}}\log d \log n + (cn^{5/6}\log{n})^{6/5} = n^{1+O(1/d)}d\log{n}$ as desired. All iterations of the loop can be done in parallel, so the number of rounds is bounded by $d + \log_2 d$ if \texttt{GetMax} is called, and by $d + 100$ if \texttt{100-Sort} is called.
\end{proof}

\subsection{Dense Selection}

Here we give two algorithms: \texttt{Select+} and \texttt{Select-}, the former of which will return a good approximation if $|L_x| > \frac{1}{10}n^{2/3}$, and the latter if $|R_x| > \frac{1}{10}n^{2/3}$. 

The idea is simple: take a large sample (size $cn^{5/6} \log{n}$), sort it with a constant approximate algorithm, and return the element in roughly the $k$-th position.

\begin{algorithm}[H]
\caption{Dense Selection}\label{alg:cap}
\begin{algorithmic}[1]
\Function{Select$\pm$}{$X, k$} 
    \State $n \gets |X|$
    \State Generate a subset $S$ of $X$ of size $c n^{5/6} \log{n}$
    \State $S \gets$ 100-Sort$(S)$
    \State \Return $S[c k\log n / n^{1/6} \pm c n^{5/12}\log n]$
\EndFunction
\end{algorithmic}
\end{algorithm}

Let $x^*$ be the item returned by \texttt{Select$\pm$}.
\begin{lemma}
    If $|L_x| > \frac{1}{10} n^{2/3}$, for any $r>0$ there exists $c$ large enough such that \texttt{Select$-$} returns a $201$-approximation with probability $> 1 - \frac{1}{n^r}$.
\end{lemma}
\begin{proof}
    It suffices to prove the actual $(k/n^{1/6} - n^{5/12})$-th smallest element of $S$ is in $L_x$, since \texttt{100-Sort} returns a $200$-approximation. The number of elements of $S$ that are $\leq x_k$ (call this $V$) follows a Hypergeometric$(n, k, cn^{5/6}\log{n})$\footnote{Similarly to before, the second argument can be $>k$, but it only makes the bounds better.} distribution. Thus, by a tail bound:
    \begin{align*}
        \pr{V \leq ck\log{n}/n^{1/6} - cn^{5/12}\log{n}} &= \pr{V \leq (k/n - n^{-5/12})cn^{5/6}\log{n}}\\
        &\leq e^{-2(n^{-5/12})^2c n^{5/6} \log{n}}\\
        &\leq n^{-2c}.
    \end{align*}
    Similarly, the number of elements of $S$ that are $< x_k - 1$ (call this $U$) follows a Hypergeometric$(n, k-\frac{1}{10} n^{2/3}, cn^{5/6}\log{n})$\footnote{Again, the second argument could be larger.} distribution. Thus, by a tail bound:
    \begin{align*}
        \pr{U \geq k/n^{1/6} - n^{5/12}} &= \pr{((k - n^{2/3}/10)/n + n^{-1/3}/10 - n^{-5/12})cn^{5/6}\log{n}} \\
        &\leq e^{-2(n^{-1/3}/10 - n^{-5/12})^2cn^{5/6}\log{n}}\\
        &< n^{-2c}.
    \end{align*}
    for $n$ sufficiently large. Thus, the probability of the actual $(k/n^{1/6} - n^{5/12})$-th smallest element is in $L_x$ is at least $1 - 2n^{-2c} > 1 - \frac{1}{n^r}$ for $c$ sufficiently large by a union bound.

\end{proof}

\begin{corollary}
    If $|R_x| > \frac{1}{10} n^{2/3}$, for any $r > 0$ there exists $c$ large enough such that \texttt{Select$+$} returns a $201$-approximation with probability $> 1 - \frac{1}{n^r}$.
\end{corollary}
\begin{proof}
    Symmetric.
\end{proof}

\begin{lemma}
    If $x^*$ is the item returned by \texttt{Select-}, for any $r > 0$ there exists $c$ sufficiently large that $x_k \geq_{200} x^*$ with probability $> 1 - \frac{1}{n^r}$.
\end{lemma}
\begin{proof}
    This is implicitly proven in the previous lemma, where we prove the position in the original array of $x^*$ is less than $k$.
\end{proof}

\begin{corollary}
    If $x^*$ is the item returned by \texttt{Select+}, for any $r > 0$ there exists $c$ sufficiently large that $x^* \geq_{200} x_k$ with probability $> 1 - \frac{1}{n^r}$
\end{corollary}
\begin{proof}
    Symmetric.
\end{proof}

\begin{theorem}
    \texttt{Select$\pm$} uses $O(n \log^{6/5} n)$ comparisons and $100$ rounds.
\end{theorem}
\begin{proof}
    All comparisons are done in \texttt{100-Sort}, so the number of comparisons is $O((n^{5/6}\log{n})^{6/5}) = O(n\log^{6/5}n)$. Since \texttt{100-Sort} takes $100$ rounds, so does \texttt{Select$\pm$}.
\end{proof}

\subsection{Combining}

\begin{algorithm}[H]
\caption{Pivot}\label{alg:cap}
\begin{algorithmic}[1]
\Function{Count}{$X, x_i$} 
    \State $c \gets 0$
    \For{$x \in X$}
        \If {$x <_c x_i$} 
            \State $c \gets c + 1$
        \EndIf
    \EndFor
    \State \Return $c$
\EndFunction
\end{algorithmic}
\end{algorithm}

\begin{algorithm}[H]
\caption{Selection}\label{alg:cap}
\begin{algorithmic}[1]
\Function{Select}{$X, k$} 
    \State $x_i \gets$ SelectSparse$(X,k)$
    \State $c_i \gets \texttt{Count}(X,x_i)$
    \If {$c_i < k$}
        \State $x_j \gets$ Select$-(X,k)$
        \If{$x_j >_c x_i$}
           \Return $x_j$
        \Else{} \Return $x_i$
        \EndIf
    \Else
        \State $x_j \gets$ Select$+(X,k)$
        \If{$x_j <_c x_i$}
            \Return $x_j$
        \Else {}
            \Return $x_i$
        \EndIf
    \EndIf
\EndFunction
\end{algorithmic}
\end{algorithm}

\begin{lemma}
    For any $r > 0$, we can choose $c$ sufficiently large that $|x_k - x^*| \leq \max(200, 1 + \min(|x_k - x_i|, |x_k - x_j|))$ with probability $> 1 - \frac{1}{n^r}$.
\end{lemma}

\begin{proof}
    By symmetry, we may assume without loss of generality that $c_i < k$. In this case, we return the item with the larger \texttt{Count}. Since we return the maximum of $x_i$ and $x_j$ (according to the comparator), we must have $\max(x_i, x_j) - 1 \leq x^* \leq \max(x_i, x_j)$. By a result from the previous section, $x_j \leq x_k + 200$ with probability $> 1 - \frac{1}{n^r}$. If $x_i$ was $> x_k + 1$, then it would compare greater than $x_k$ and everything before it, contradicting $c_i < k$. Thus, $x_i \leq x_k + 1$. It follows that with probability $> 1 - \frac{1}{n^r}$, $\max(x_i, x_j) \leq x_k + 200$. Thus, if either $x_i > x_k$ or $x_j > x_k$, $|x^* - x_k| \leq 200$ as desired. Otherwise, if both $x_i \leq x_k$ and $x_j \leq x_k$, then $|x_k - x^*| = x_k - x^* \leq x_k - (\max(x_i, x_j) - 1) = \min(|x_k - x_i|, |x_k - x_j|) + 1$ as desired.
\end{proof}

\begin{theorem}
    For $r > 0$ there exists $c$ sufficiently large that \texttt{Select} returns a $202$-approximate $k$-selection with probability $> 1 - \frac{1}{n^r}$.
\end{theorem}
\begin{proof}
    We consider cases based on the sizes of $L_x$ and $R_x$:

    If $|L_x| \leq \frac{1}{10}n^{2/3}$ and $|R_x| \leq \frac{1}{10}n^{2/3}$, then we have $x_i \geq_{200} x_k$ and $x_k \geq_{200} x_i$ with probability $> 1 - \frac{2}{n^{r+1}}$, in which case we have $|x_k - x_i| \leq 200$. By the previous lemma it follows that $|x^* - x_k| \leq 201$ with probability $> 1-\frac{3}{n^{r+1}}$, so we return a $201$-approximate $k$-selection with probability $> 1-\frac{3}{n^{r+1}} > 1-\frac{1}{n^r}$ as desired.

    If $|L_x| > \frac{1}{10}n^{2/3}$ and $|R_x| > \frac{1}{10}n^{2/3}$, then $|x_j - x_k| \leq 201$ with probability $> 1-\frac{1}{n^{r+1}}$. Thus, by the previous lemma, $|x^*-x_k| \leq 202$ with probability $> 1-\frac{2}{n^{r+1}}$. It follows that $x^*$ is a $202$-approximate $k$-selection with probability $> 1-\frac{2}{n^{r+1}}>1-\frac{1}{n^r}$ as desired.
    
    If $|L_x| \leq \frac{1}{10} n^{2/3}$ and $|R_x| > \frac{1}{10}n^{2/3}$, we have $x_i \geq_{200} x_k$ with probability $> \frac{1}{n^{r+1}}$. Thus, either $x_i \leq x_k + 1$ in which case we have $|x_i - x_k| \leq 200$, or $x_i > x_k + 1$ in which case $x_j$ must come from \texttt{Select+} and thus $|x_j - x_k| \leq 201$ with probability $> \frac{1}{n^{r+1}}$. By the previous lemma, it thus follows that $|x^* - x_k| \leq 202$ with probability $>\frac{3}{n^{r+1}}$. It follows that $x^*$ is a $202$-approximate $k$-selection with probability $>1-\frac{3}{n^{r+1}}>1-\frac{1}{n^r}$ as desired.

    The case where $|L_x| > \frac{1}{10}n^{2/3}$ and $|R_x| \leq \frac{1}{10}n^{2/3}$ is symmetric.
\end{proof}

\begin{theorem}
    \texttt{Select} takes $n^{1+O(1/d)}d \log{n}$ comparisons and $d + 102 + \min(100, \log_2 d)$ rounds. 
\end{theorem}
\begin{proof}
    All comparisons are done in \texttt{SelectSparse}, \texttt{Select$\pm$} and \texttt{Count}. The number of comparisons done by the two calls to \texttt{Count} is bounded by $2n$. Thus, the total number of comparisons is bounded by $n^{1+O(1/d)} d \log{n} + n\log^{6/5}n + 2n = n^{1+O(1/d)}d\log{n}$. Each call to \texttt{Count} takes one round, so the total number of rounds is bounded by $d + \min(100,\log_2 d) + 100 + 2 = d + 102 + \min(100, \log_2 d)$.
\end{proof}

Theorem~\ref{theorem:SelectR} follows.

\section{Open Problems}

\begin{itemize}
    \item[--] Is there an algorithm to find a $3$-approximate sorting or selection with high probability in $\widetilde{O}(n)$ time?
    \item[--] Is there an algorithm to find a constant-approximate sorting with high probability in $O(n \log n)$ time?
    \item[--] Is there an algorithm to find a constant-approximate selection with high probability in $O(n)$ time?
    \item[--] Can we improve the lower or upper bounds for $k$-approximate sorting and selection in $d$ rounds?
\end{itemize}

\section*{Acknowledgements}

I would like to thank Gautam Kamath for introducing me to this problem, advising me throughout this process, and giving feedback on earlier drafts of this paper. This would not have been possible without his help. I would also like to thank Richard Peng for pointing me in the direction of Gautam, and Yousof Hosny for helpful discussions. Additionally, I would like to thank Guandi Gani for pointing out some minor errors in a previous version of this paper.


\appendix

\bibliographystyle{alpha}
\bibliography{biblio}

\end{document}